\documentclass[runningheads]{llncs}
\usepackage[T1]{fontenc}
\usepackage{graphicx}
\usepackage{multirow}
\usepackage{comment}
\usepackage{subfig}
\usepackage{float}
\def\ninept{\def\baselinestretch{.95}\let\normalsize\small\normalsize}
\ninept
\begin{document}
	\title{Exploring Effective Fusion Algorithms for Speech Based Self-Supervised Learning Models}
	\titlerunning{Exploring Effective Fusion Algorithms for Speech Based SSL Models}
	%
	\author{Changli Tang$^\dagger$\inst{1}\orcidID{0000-0002-2009-3078},
		Yujin Wang$^\dagger$\inst{1}\orcidID{0000-0001-6188-5672},
		Xie Chen$^\ddagger$\inst{2}\orcidID{0000-0001-7423-617X}, Wei-Qiang Zhang$^\ddagger$\inst{1} 
		\thanks{$^{\dagger}$ equal contribution, $^{\ddagger}$ corresponding authors}}
	\authorrunning{C. Tang et al.}
	%
	\institute{
	$^1$Department of Electronic Engineering, Tsinghua University, Beijing, China \\
	$^2$MoE Key Lab of Artificial Intelligence, AI Institute \\ X-LANCE Lab, Department of Computer Science and Engineering \\ Shanghai Jiao Tong University, Shanghai, China \\
		\email{\{tcl20, yujin-wa20\}@mails.tsinghua.edu.cn} \\
	}
	\maketitle              
\begin{abstract}
Self-supervised learning (SSL) has achieved great success in various areas including speech processing. Recently, it is proven that speech based SSL models are able to extract superior universal representations on a range of downstream tasks compared to traditional hand-craft feature (e.g. FBank, MFCC) in the SUPERB benchmark. However, different types of SSL models might exhibit distinct strengths on different downstream tasks. In order to better utilize the potential power of SSL models, in this work, we explore the effective fusion  on multiple SSL models. A series of model fusion algorithms are investigated and compared by combining two types of SSL models, Hubert and Data2vec, on two representative tasks from SUPERB benchmark, which are speaker identification (SID) and automatic speech recognition (ASR) tasks. The experimental results demonstrate that our proposed fusion algorithms can further boost the individual model significantly.
		
\end{abstract}

\section{Introduction}
In recent years, self-supervised learning (SSL) has made great progress in speech representation learning~\cite{review}. The general idea of SSL is to reconstruct or predict itself based on its context information, which allows the model to learn the underlying structure information effectively in an unsupervised way. As a result, the SSL model can be pre-trained on oceans of unlabelled speech data, and then fine-tuned with a small amount of transcribed speech on the specific downstream task to achieve significant performance improvement. To data, a series of successful speech based SSL speech models have been developed, such as Wav2vec 2.0 \cite{wav2vec2}, HuBERT \cite{HuBERT}, WavLM \cite{wavlm} and Data2vec \cite{data2vec}.

In most SSL studies, speech recognition is considered as the main task for performance evaluation. More recently, the SUPERB benchmark \cite{SUPERB} was proposed and built to validate whether the well-trained SSL models are capable of extracting superior universal feature on a range of downstream tasks, instead of merely on the speech recognition task. This provides a flexible and straightforward way to analysis the strength and weakness of different SSL models on different tasks. 
	
The public experimental results on the SUPERB leaderboard have shown that SSL models (e.g. HuBERT, Data2vec) exhibit complementary performances on various downstream tasks. As shown in Table \ref{tab:1}, we can conclude that with the similar model size, HuBERT performs better on speaker-related tasks, while Data2vec is superior on content-related tasks. Motivated by this, in this work, we investigate on the effective model fusion of these SSL models. A series of model fusion algorithms are proposed and compared, with a view to generating richer representations and combining the strengths of different pre-trained models. According to our preliminary experimental results, by applying effective model fusion algorithms, the performance on the specific downstream tasks can be further improved significantly over the individual model. It is worth mentioning that the combination of Hubert and Data2vec achieves the SOTA performance in the ASR track of SUPERB leaderboard \footnote{https://superbbenchmark.org/leaderboard}, which yields 6.5\% relative WER reduction over Data2vec Large model, and 13.2\% relative WER reduction over Hubert Large model. 
	
\section{Related Works}
\subsection{Speech based Self-Supervised Learning}
In recent years, several speech based SSL models have been developed and proven to be effective for downstream tasks. In this section, we give a brief overview of two representative SSL models, Hubert \cite{HuBERT} and Data2vec \cite{wav2vec2}. These two models are also chosen for model fusion in the following sections. In Hubert, K-Means clustering is applied to cluster the speech frames into a specified number of classes. The input of the K-Means can be either the standard MFCC feature or the hidden representation of a well-trained Hubert model. As a result, each frame is assigned to one class. And the model was constructed to predict its class id for each frame with standard cross-entropy loss. It is noting that the current input frame for prediction will be masked to encourage the utilization of context information and avoid  information leakage. The Hubert model can be viewed as a classification task. In contrast, Data2vec attempts to construct a regression task for self-supervised learning. To this end, two parallel models are maintained during the training of Data2vec, which are teacher and student models. The teacher model is obtained by apply the exponential moving average (EMA) technique on the student model, which could be viewed as a delayed version of the student model. The student model takes the masked input, while the teacher consumes the complete input without masking.  The L2 distance between the outputs of the teacher and student model is minimized during training. 

In literature, Data2vec is reported to yield better WER performance compared to Hubert in speech recognition. However, the conclusion turns to be mixed when considering other downstream tasks. We will explain this in more detail in the following sections.
	
\subsection{SUPERB Benchmark}
SUPERB is a popular benchmark for self-supervised pretrained models which aims to measure their performance on different downstream tasks. When a pretrained model is fine-tuning on a specified downstream task, its feature extractor (e.g. CNN models with waveform input) and Transformer encoder will be frozen, and the outputs of each Transformer layer will be summed with learnable weights. This weighted-sum output is finally fed into a task-specific downstream expert.
	
\begin{figure}[htbp]
		\centering
		\caption{Weights analysis of HuBERT and Data2vec in different downstream tasks in SUPERB benchmark}
		\subfloat[HuBERT]{\includegraphics[width=0.5\textwidth]{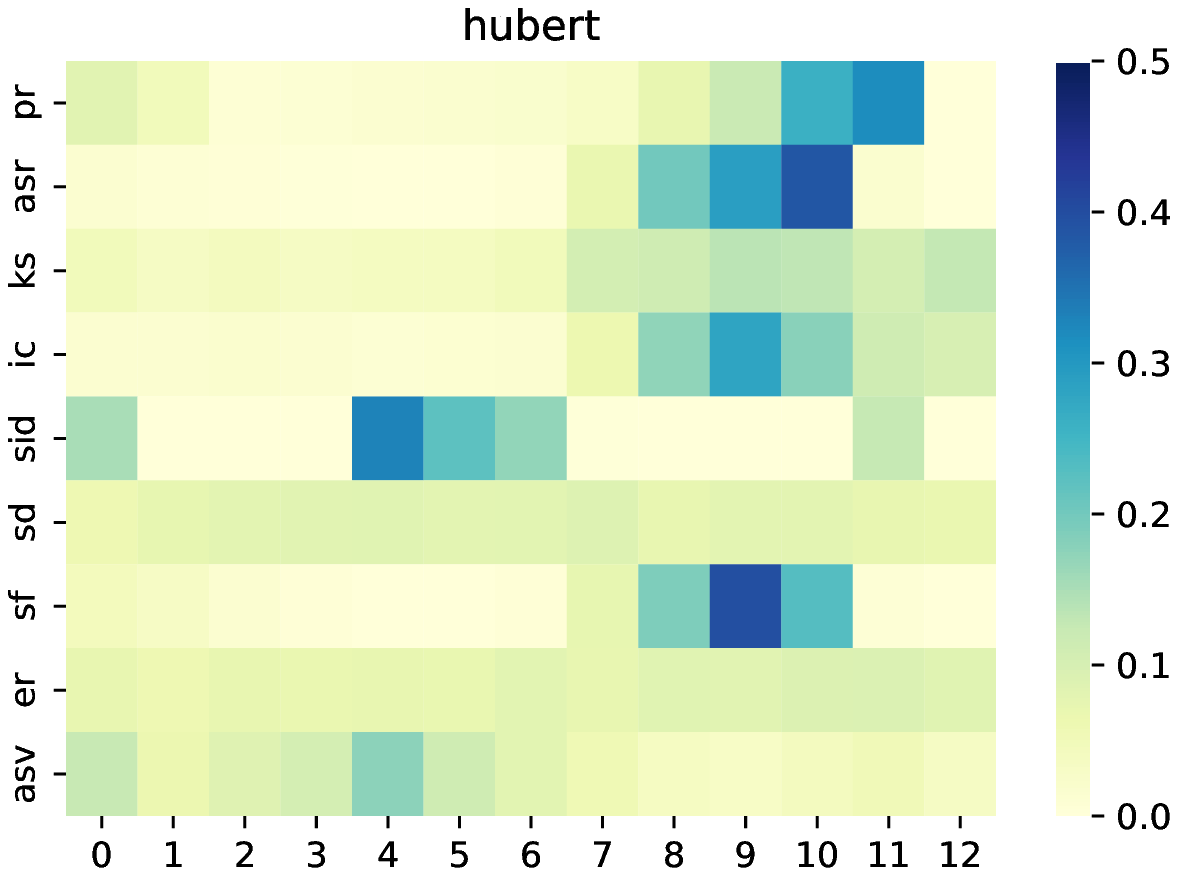}}
		\subfloat[Data2vec]{\includegraphics[width=0.5\textwidth]{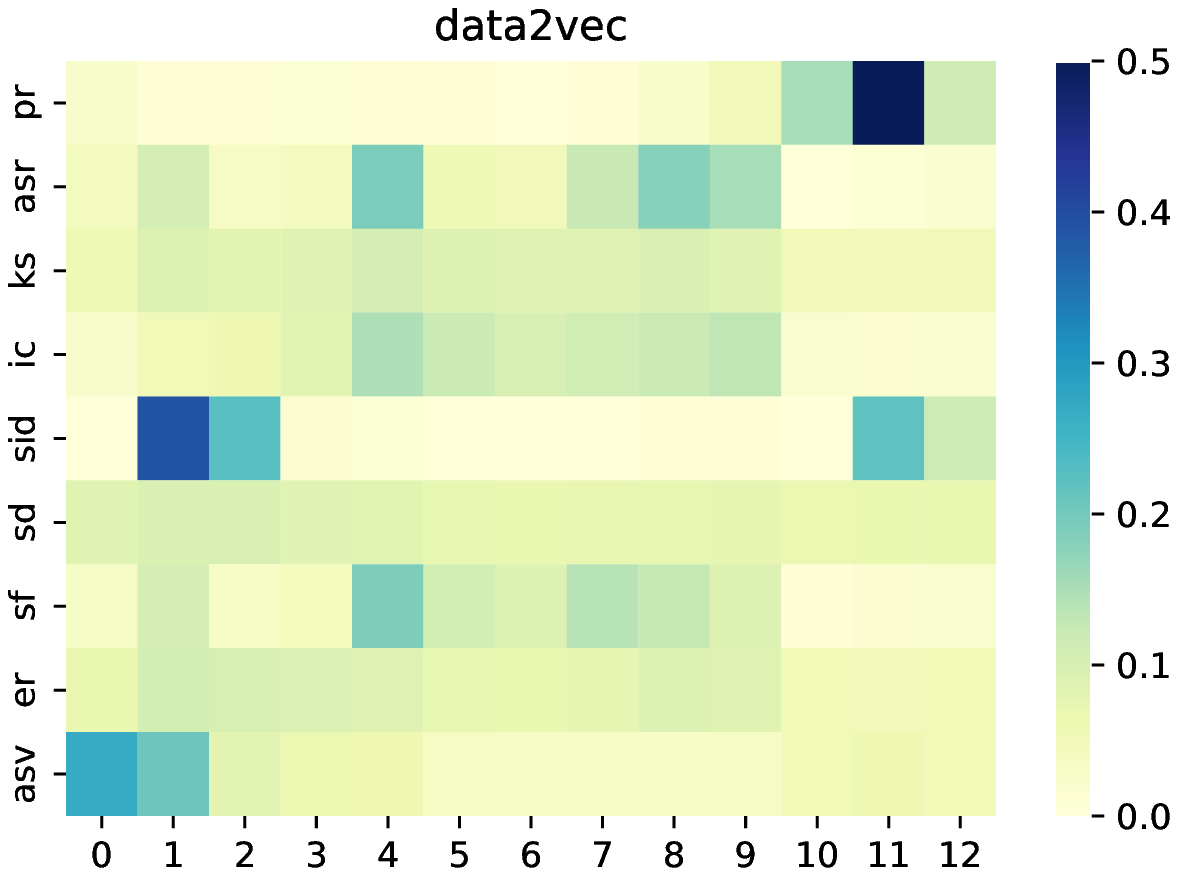}}
		\label{fig:weight}
	\end{figure}

In this work, Data2vec base and large models are evaluated on various downstream tasks on SUPERB \footnote{The results of Data2vec is also submitted and updated in the SUPERB leaderboard:  https://superbbenchmark.org/leaderboard}, including Speaker Identification (SID), Automatic Speaker Verification (ASV), Speaker Diarization (SD), Phoneme Recognition (PR), Automatic Speech Recognition (ASR), Keyword Spotting (KS), Query by Example Spoken Term Detection (QbE), Intent Classification (IC), Slot Filling (SF), and Emotion Recognition (ER). These downstream tasks are associated with different information level lied in speech, and they can be simply classified into four categories: speaker, content, semantics and paralinguistics. Table~\ref{tab1} gives our results on Data2vec and compared with the public numbers on Hubert. It can be seen that Data2vec presents strong capabilities on content-related tasks such as ASR, while yields poor performance on speaker-related tasks such as SID. The weight distribution of each layer for Hubert and Data2vec on SUPERB benchmark is also plotted in Figure \ref{fig:weight}.
	
Our model fusion is mainly based on Data2vec and HuBERT, as they are two most representative SSL models and are quite complementary on different downstream tasks. We hope that the combination of Data2vec and HuBERT models can produce an all-powerful model on SUPERB. 
	
    \begin{table}
		\caption{Results of HuBERT and Data2vec in different downstream tasks in SUPERB benchmark.}\label{tab1}
		\resizebox{\textwidth}{15mm}{
			\begin{tabular}{c|c|c||c|c|c||c|c|c|c||c|c|c||c}
				\hline 
				\multirow{3}*{Method} & \multirow{3}*{\#Params} & \multirow{3}*{Corpus} & \multicolumn{3}{|c||}{Speaker} & \multicolumn{4}{|c||}{Content} & \multicolumn{3}{|c||}{Semantics} & ParaL  \\
				\cline{4-14} & & & SID & ASV & SD & PR & ASR & KS & QbE & IC & \multicolumn{2}{c||}{SF} & ER \\
				\cline{4-14} & & & ACC $\uparrow$ & EER $\downarrow$ & DER  $\downarrow$ & PER  $\downarrow$ & WER  $\downarrow$             & ACC $\uparrow$ & MTWV $\uparrow$ & Acc $\uparrow$ & F1 $\uparrow$  & CER $\downarrow$ & Acc $\uparrow$ \\
				\hline
				\hline 
				Data2vec Base & 93.75M & LS 960 hr & 70.21 & 5.77 & 6.67 & \textbf{4.69} & \textbf{4.94} & \textbf{96.56} & 0.0576 & 97.63 & \textbf{88.59} & 25.27 & \textbf{66.27} \\
				HuBERT Base & 94.70M & LS 960 hr & \textbf{81.42} & \textbf{5.11} & \textbf{5.88}  & 5.41 & 6.42 & 96.30 & \textbf{0.0736} & \textbf{98.34} & 88.53 & \textbf{25.20} & 64.92 \\
				\hline
				\hline         
				Data2vec Large & 314.3M & LL 60k hr & 76.77 & \textbf{5.73} & \textbf{5.53} & 3.60 & \textbf{3.36} & \textbf{96.75} & \textbf{0.0628} & 98.31 & \textbf{90.98} & 22.16 & 66.31 \\
				HuBERT Large & 316.6M & LL 60k hr & \textbf{90.33}  & 5.98 & 5.75 & \textbf{3.53} & 3.62 & 95.29 & 0.0353 & \textbf{98.76} & 89.81          & \textbf{21.76}   & \textbf{67.62} \\
				\hline
			\end{tabular}
		}
	\label{tab:1}
	\end{table}
	
\subsection{Model Fusion}
As Figure~\ref{fig:weight} shows, the allocation of hidden layer weights on different downstream tasks behave quite differently between HuBERT and Data2vec, which provides a theoretical basis for our model fusion.
	
\cite{model_fusion_india} proposed a model fusion framework for different pre-trained models. The SSL models are frozen and the hidden features generated by the last layer from different SSL models are weighted and summed. However, previous analyses \cite{layer} showed that the semantic information of the model was not mainly distributed in the last layer, so only using representation from the last hidden layer might not be able to extract the sufficient semantic information.

\section{Methods}
\begin{figure}[htbp]
		\centering
		\subfloat[Naive feature-level fusion ]{\includegraphics[width=.48\textwidth]{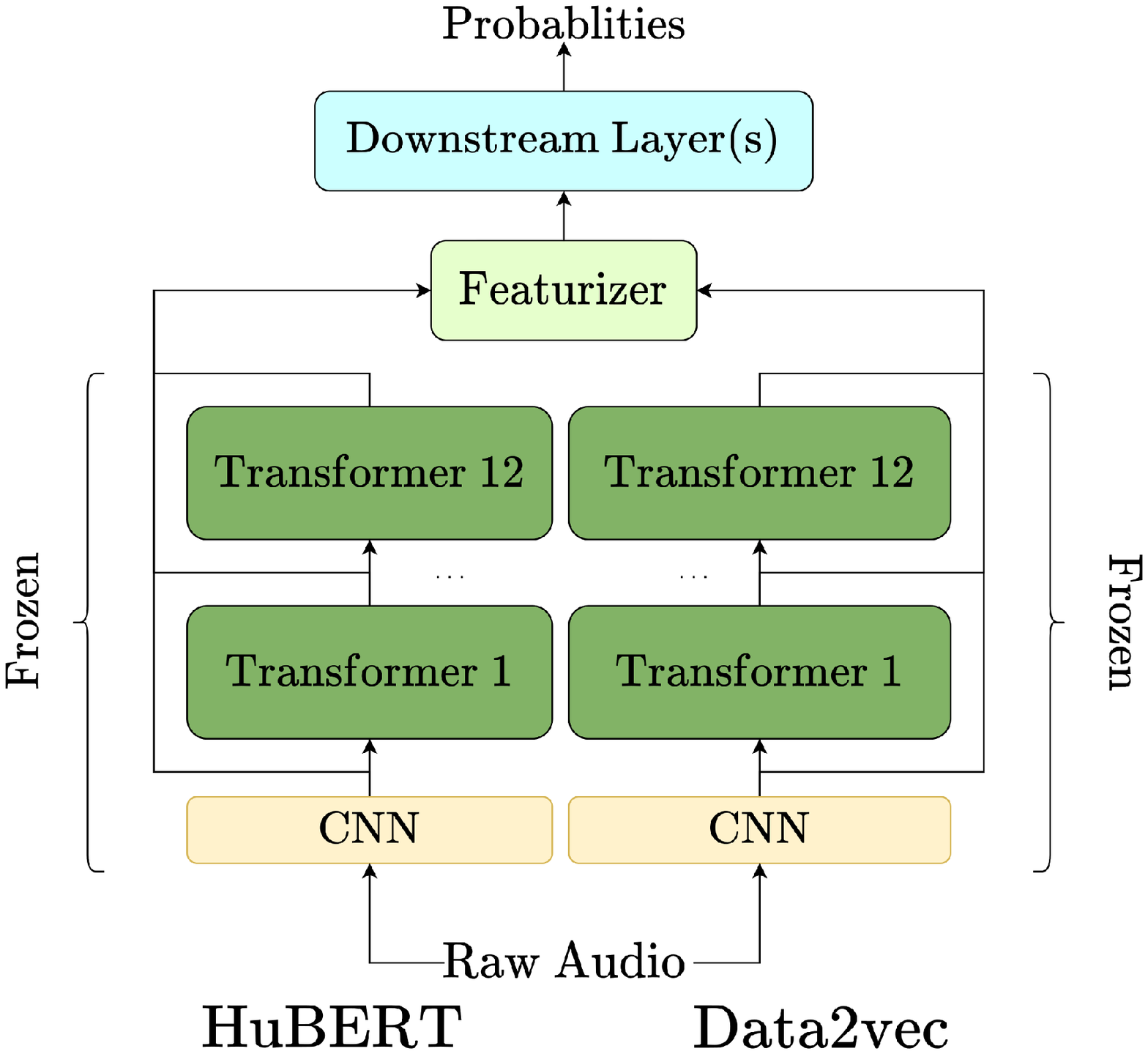}}\hspace{5pt}
		\subfloat[Structured feature-level fusion ]{\includegraphics[width=.48\textwidth]{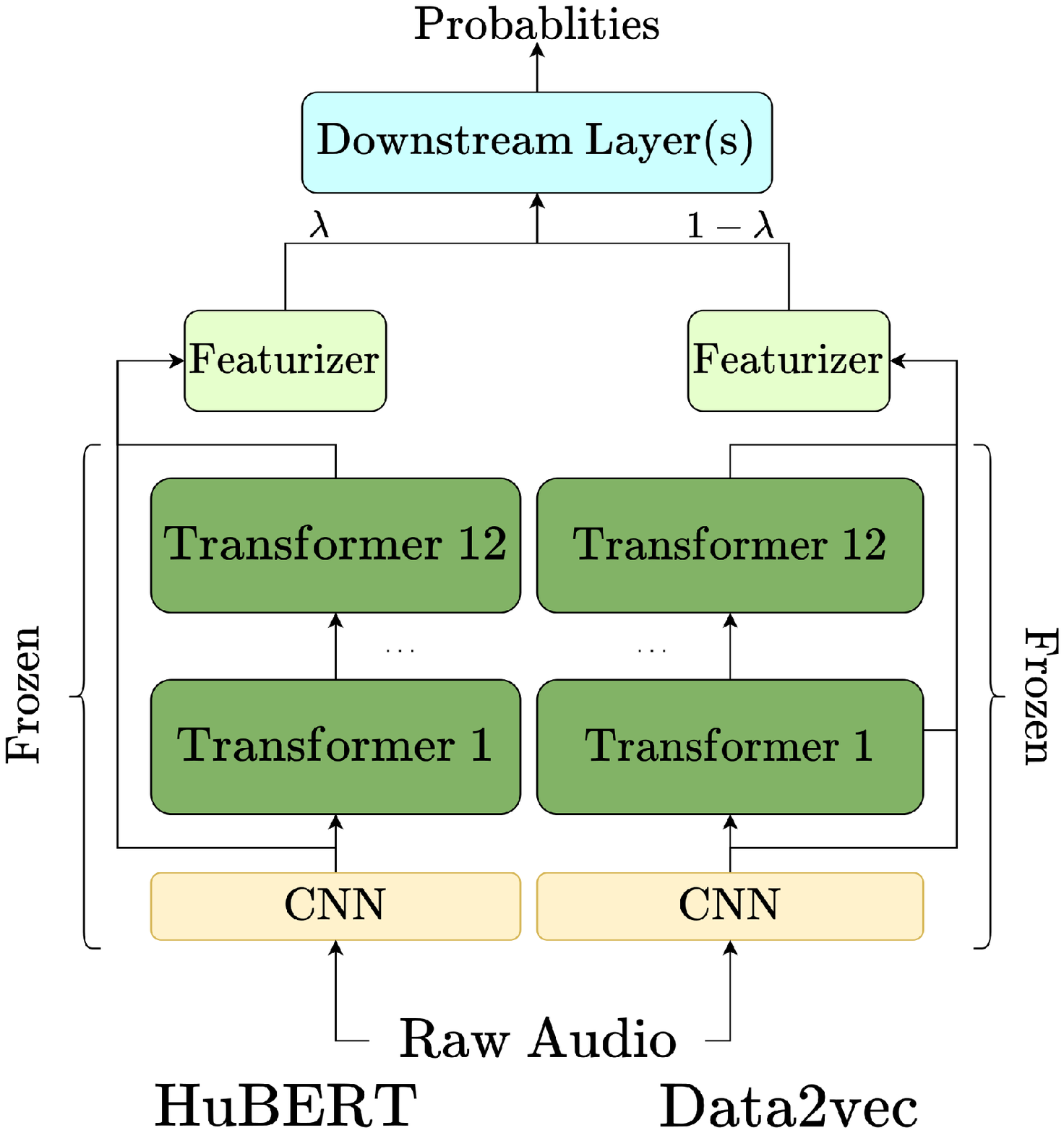}}\\
		\subfloat[Probability-level fusion with shared head ]{\includegraphics[width=.48\textwidth]{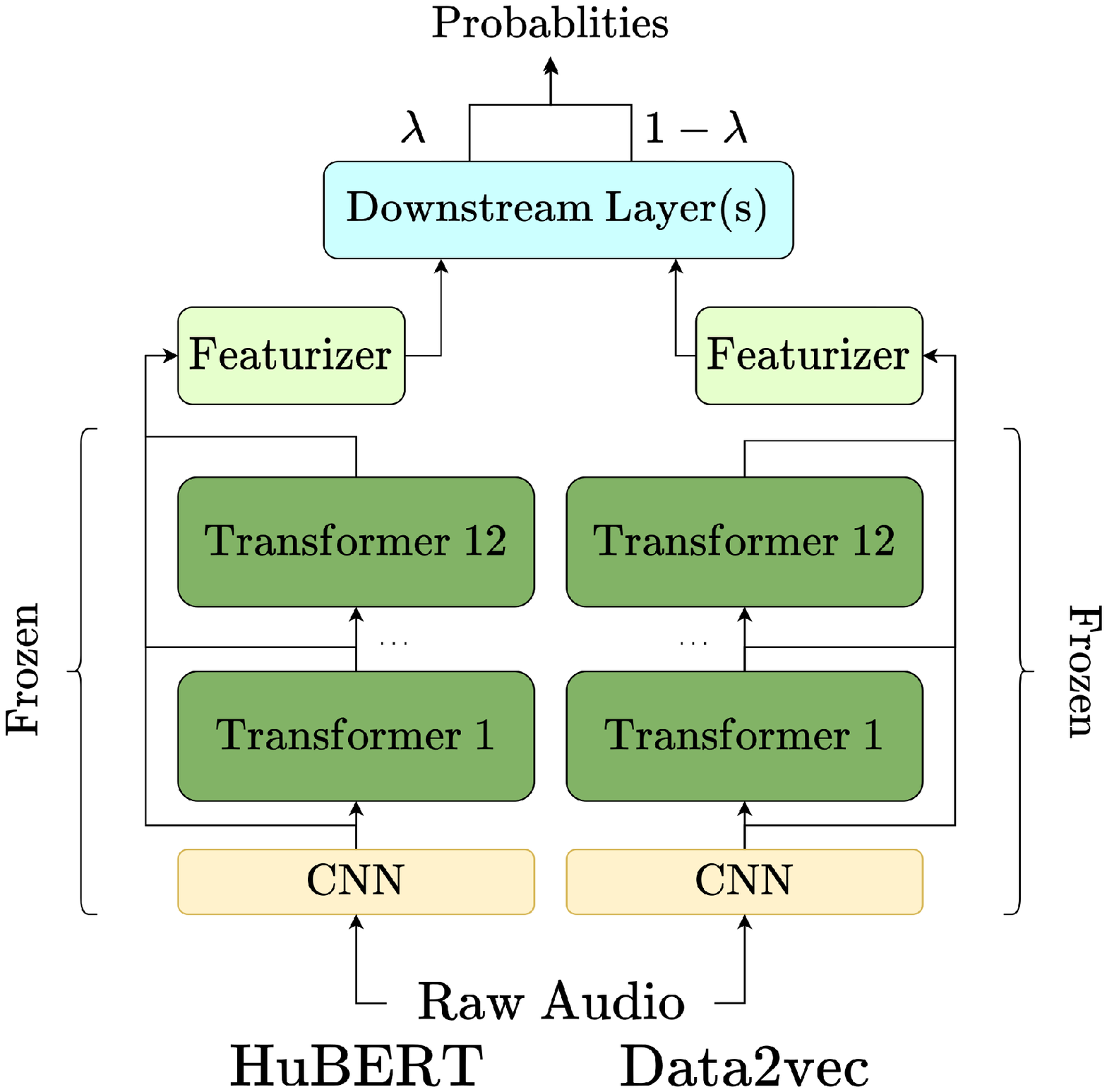}}\hspace{5pt}
		\subfloat[Probability-level fusion with individual heads ]{\includegraphics[width=.48\textwidth]{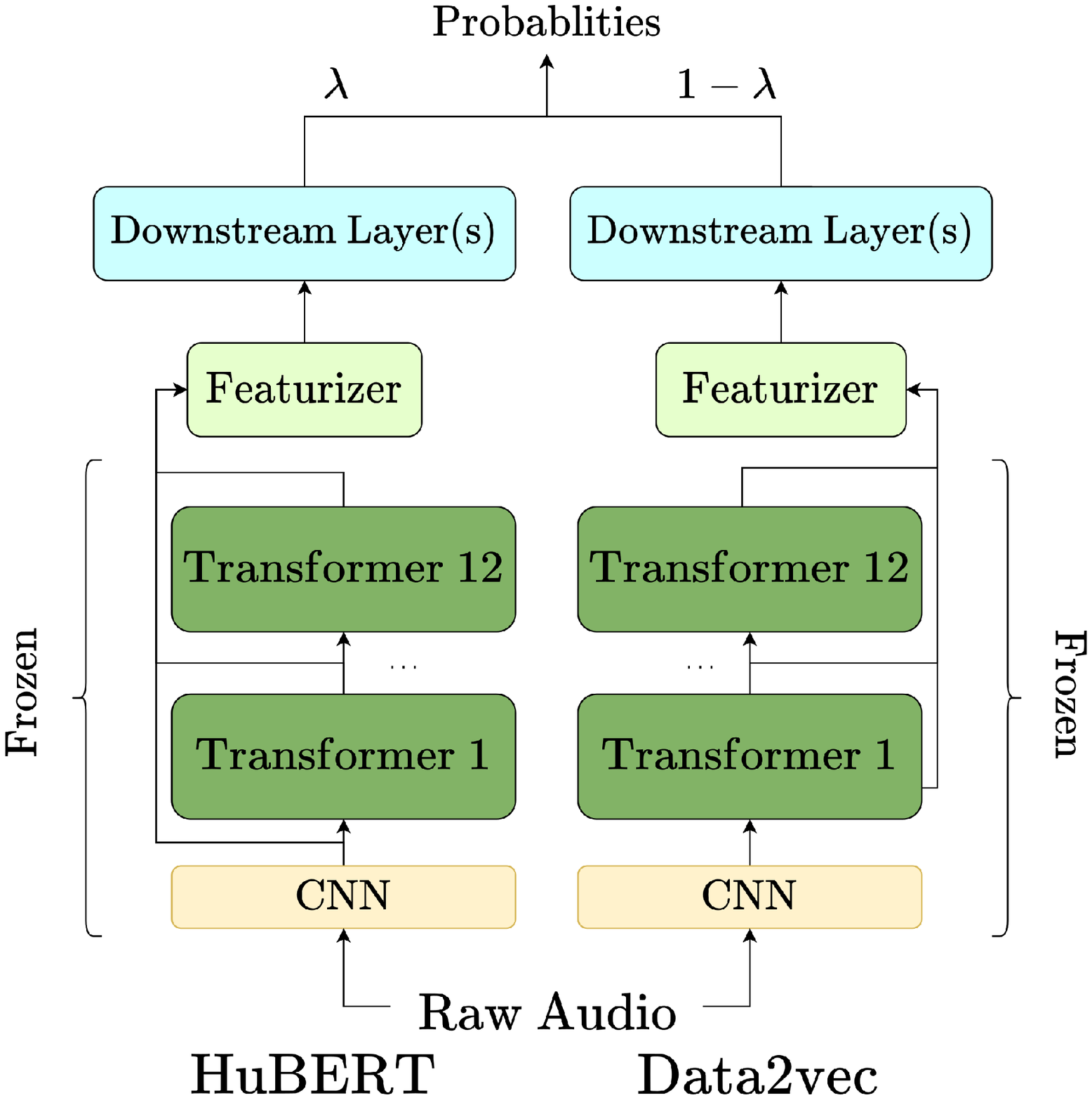}}
		\caption{\\(a): The hidden layer outputs of the two models directly weighted by the same featurizer. \\ (b): The respective hidden layer outputs in each model are first weighted and summed, and then the outputs of each model are weighted and fed to the downstream model. \\ (c): The probabilities of the two models are generated using a shared downstream models respectively, and later weighted. \\ (d): The probabilities for the two models are generated separately using separate downstream models and later weighted.}
		\label{framework}
	\end{figure}

In this paper, we propose and compare four fusion methods for multiple self-supervised models: two feature-level fusions, and two probability-level fusions, as shown in Figure \ref{framework}. The stage of model fusion of these methods is gradually extended backwards in turn.
	
Let function $\mathcal{F}$ denote the downstream model, $m$ be the number of models to be fused, $l$ be the number of layers of each model, $w_{ij}$, $\textbf{h}_{ij}$ denote the weights and hidden vectors of the $j^{th}$ layer of features of the $i^{th}$ model respectively. The output of the downstream model is considered to be the probability against the task label.
	
The first way is to simply fuse hidden representations from different layers in all models. With a large featurizer, the hidden vectors of each layer of the different models are directly weighted-summed and then fed into the downstream head.
	\begin{equation}\label{eq1}
		{\rm prob} = \mathcal{F}(\sum_{i=1}^{m}\sum_{j=0}^{l}w_{ij}\textbf{h}_{ij}) \quad \sum_i\sum_jw_{ij}=1
	\end{equation}
	
The second way is to fuse the hidden representations of different models in a hierarchical way. First, we weighted-sum the outputs of different layers of each SSL model. We then apply a second weighted sum of these outputs from the first step, to compute the input for the downstream task-specific model.
	\begin{equation}\label{eq2}
		{\rm prob} = \mathcal{F}(\sum_{i=1}^{m}p_i\sum_{j = 0}^{l}w_{ij}\textbf{h}_{ij}) \quad \sum_ip_i=1 \quad \forall i, \sum_jw_{ij}=1 
	\end{equation}
	
The third way is to fuse the probability distributions of different upstream models. For each SSL model, we weighted-sum outputs of different layers and feed the result into the downstream head. The output of the downstream head forms a probability distribution for task labels. We combine the probability distributions of different models and use the fused probability distribution for inference.
	\begin{equation}\label{eq3}
		{\rm prob} = \sum_{i=1}^mp_i \mathcal{F}(\sum_{j=0}^lw_{ij}\textbf{h}_{ij}) \quad \sum_ip_i=1 \quad \forall i, \sum_jw_{ij}=1
	\end{equation}
	
The fourth way is similar to the third, but we use multiple downstream heads here. Each upstream model has its own corresponding downstream model. The probability distributions generated by the different downstream models will be fused and the fused distributions will be used for inference on the actual labels.
	\begin{equation}\label{eq4}
		{\rm prob} = \sum_{i=1}^mp_i \mathcal{F}_i(\sum_{j=0}^lw_{ij}\textbf{h}_{ij}) \quad \sum_ip_i=1 \quad \forall i, \sum_jw_{ij}=1
	\end{equation}
	
\section{Experiments and Analysis}
In this section, the above-mentioned four model fusion approaches are investigated 
and compared on the SUPERB benchmark. Following the constrained track in SUPERB, the SSL upstream models are frozen during training. Only the downstream head and the weights of hidden vectors are learnable. We choose Data2vec and HuBERT as the candidate models for model fusion, hoping to bring their strengths together. We focus primarily on the SID and ASR tasks, as two representative speaker-related and content-related tasks.
	
In the first experiment, the are only two models to be fused, $m = 2$, and the Data2vec and HuBERT base models both have 12 transformer layers, $l = 12$.
For simplicity, in both feature-level and probability-level model fusion, we fix the model-level interpolation weight $p_i$ in Eqs. \ref{eq2}, \ref{eq3} and \ref{eq4} to $0.5$, i.e.
	$$
	p_1=p_2=0.5
	$$
	
Four model fusion methods are applied and the results are shown in Table \ref{tab:addlabel}. For the ASR task, we followed the default settings of SUPERB, fine-tuning 200,000 steps with a learning rate 1e-4. For the SID task, we similarly fine-tuned 200,000 steps. However, due to the SID task is sensitive to the learning rate, we conducted a search for the learning rate and finally set it to 0.1.
	
	\begin{table}[htbp]
		\centering
		\caption{Fusion results of Data2vec and HuBERT on the SID and ASR tasks in SUPERB benchmark}
		\begin{tabular}{c|c|c}
			\hline
			\textbf{Fusion level} & {\textbf{SID (acc)}} & {\textbf{ASR (w/o LM wer)}} \\
			\hline
			HuBERT base & 81.42 & 6.42 \\
			\hline
			Data2vec base & 70.21 & 4.94 \\
			\hline
			\hline
			Naive feature-level & 48.94 & 5.03 \\
			\hline
			Structured feature-level & 78.98 & {\bf 4.62} \\
			\hline
			Probability-level w/ shared head & 80.14 & 5.14 \\
			\hline
			Probability-level w/ individual heads & {\bf 86.04} & 5.04 \\
			\hline
		\end{tabular}%
		\label{tab:addlabel}%
	\end{table}	
	
From Table \ref{tab:addlabel}, we can see that naively fusing features of different models performs poorly on SID task. The accuracy of direct fusion is even much lower than that of Data2vec, suggesting that directly blending features from both models is likely to significantly degrade performance on speaker-related tasks. 
	
Interestingly, a structured feature-level fusion can largely slow down the performance degradation on the SID task, while this also helps with the ASR task. This structured fusion method outperforms naive fusion, indicating to some extent that the hidden states within the models are somewhat linked, while the hidden states between the models are more different. Therefore, if the features are fused directly, this naive fusion tends to confuse the information as the hidden states of the different models are quite different. However, 
when the fusion is first conducted within the same model and then fused across model, the information within the same model can be better integrated, and so as to improve the performance in downstream tasks. 

These two probability-level fusions are quite similar, except that where they adopt a shared or separate downstream heads. Similar to the fusion on feature-level, this probabilistic fusion first integrates the model's own information, so the result of SID is not as bad as if the features were fused directly. It is important to note that although the fusion is applied on probability level, these two models jointly utilise the same downstream head and are jointly influenced by the same downstream head during forward propagation.
	
The fusion of each model with its own downstream head performs best on SID task. Because the downstream model no longer shares, information from different models is  combined after the computation of probabilities. Therefore, as can be seen from the results above, the closer the stage of information exchange is to the label of the task, the more effective the fusion will be for SID tasks.
	
For ASR task, the structured feature-level fusion of Data2vec and Hubert base model is able to largely improve the performance. The next experiment is to validate the effectiveness of this structured feature-level fusion on large models. Table \ref{tab:asrlarge} shows that this fusion approach can improve WER performance consistently for model fusion between large models. It is worth mentioning that Data2vec large model is the current SOTA on the ASR task in SUPERB leaderboard \footnote{https://superbbenchmark.org/leaderboard}, the fusion of HuBERT and Data2vec further reduce the WER and achieve a new SOTA for this task. 
	
	\begin{table}[htbp]
		\centering
		\caption{Large models fusion results of Data2vec and HuBERT}
		\begin{tabular}{c|c}
			\hline
			\textbf{Fusion level} & {\textbf{ASR (w/o LM wer)}} \\
			\hline
			HuBERT large & 3.62 \\
			\hline
			Data2vec large & 3.36 \\
			\hline
			\hline
			Structured feature-level fusion & 3.14 \\
			\hline
		\end{tabular}%
		\label{tab:asrlarge}%
	\end{table}	
	
\section{Conclusion and Future Work}
In order to improve the comprehensive performance of the models and to integrate the advantages of different models, we propose a series of model fusion methods and test them under several representative SUPERB downstream tasks. Experiments show that for speaker-related task like SID, simply fusing models on feature perform poorly. However, the later the stage of information exchange between different models, the better the performance may be. And for content-related work like ASR, the structured feature-level fusion is able to improve the performance significantly.

In our future work, we will further investigate the effective fusion algorithm for multiple SSL models, to extract better universal feature over any single SSL model in different downstream tasks. In addition, we will also investigate fusion on more than two models to explore more potential for SSL models.
	
\subsubsection{Acknowledgements} This work was supported by the National Natural Science Foundation of China under Grant No. U1836219, No. 62276153 and No. 6220070337. We extend our sincere gratitude to them.
	%
	%
	%
\bibliography{ref}

\end{document}